\documentclass[journal]{IEEEtran}
\usepackage[final]{graphicx}
\usepackage{dcolumn}
\usepackage{bm}
\usepackage{eso-pic}
\usepackage{rotating}
\usepackage{amssymb, bm}
\usepackage{color}
\usepackage{pdfsync}
\usepackage[sort&compress,numbers,square,comma]{natbib}
\usepackage{lipsum}
\newcommand{\botharrowp}{
				\begin{turn}{-45}
								\raisebox{5pt}{\ensuremath{\text{\tiny$\leftrightarrow$}}}
        \end{turn}
}
\newcommand{\botharrowm}{
				\begin{turn}{45}
                \raisebox{-3pt}{\ensuremath{\text{\tiny$\leftrightarrow$}}}
        \end{turn}
}
\newcommand{\botharrowpnorm}{
				\begin{turn}{-45}
								\raisebox{5pt}{\ensuremath{\text{\small$\leftrightarrow$}}}
        \end{turn}
}
\newcommand{\botharrowmnorm}{
				\begin{turn}{45}
                \raisebox{-3pt}{\ensuremath{\text{\small$\leftrightarrow$}}}
        \end{turn}
}

\newcommand{\blk}{\color{black}}

\definecolor{dark-blue}{rgb}{0,0,1}
\usepackage[%
linkcolor=dark-blue,
citecolor=dark-blue,
urlcolor=dark-blue,
colorlinks=true,
]{hyperref}

\ifCLASSINFOpdf
\else
\fi
\usepackage[cmex10]{amsmath}
\definecolor{vanya}{RGB}{0,153,51}

\begin{document}

\title{Stealth technology-based Terahertz frequency-domain ellipsometry instrumentation}

\author{Philipp~K\"{u}hne, Vallery~Stanishev, Nerijus~Armakavicius, Mathias~Schubert, and~Vanya~Darakchieva,
\thanks{P.~K\"{u}hne, V.~Stanishev, N.~Armakavicius, M.~Schubert and V.~Darakchieva are with the Terahertz Materials Analysis Center (THeMAC), Department of Physics, Chemistry and Biology, IFM, Link\"{o}ping University, Link\"{o}ping S-58183 SE, Sweden. M.~Schubert is also with the Department of Electrical and Computer Engineering, and the Center for Nanohybrid Functional Materials at University of Nebraska-Lincoln, Lincoln, Nebraska 68588, U.S.A.}
\thanks{Manuscript received January 1, 1800; revised January 2, 1800.}}

\markboth{IEEE TRANSACTIONS ON TERAHERTZ SCIENCE AND TECHNOLOGY}%
{Shell \MakeLowercase{\textit{et al.}}: THz TDS Ellipsometry}

\maketitle
\begin{abstract}
We present a terahertz (THz) frequency-domain spectroscopic (FDS) ellipsometer design which suppresses formation of standing waves by use of stealth technology approaches. The strategy to suppress standing waves consists of three elements \textit{geometry}, \textit{coating} and \textit{modulation}. The instrument is based on the rotating analyzer ellipsometer principle and can incorporate various sample compartments, such as a superconducting magnet, \textit{in-situ} gas cells or resonant sample cavities, for example. A backward wave oscillator and three detectors are employed, which permit operation in the spectral range of 0.1--1~THz (3.3--33~cm$^{-1}$ or 0.4--4~meV). The THz frequency-domain ellipsometer allows for standard and generalized ellipsometry at variable angles of incidence in both reflection and transmission configurations. 
The methods used to suppress standing waves and strategies for an accurate frequency calibration are presented. Experimental results from dielectric constant determination in anisotropic materials, and free charge carrier determination in  optical Hall effect, resonant-cavity enhanced optical Hall effect and \textit{in-situ} optical Hall effect experiments are discussed. Examples include silicon and sapphire optical constants, free charge carrier properties of two dimensional electron gas in group-III nitride high electron mobility transistor structure, and ambient effects on free electron mobility and density in epitaxial graphene.
\end{abstract}


\IEEEpeerreviewmaketitle

\section{Introduction}
%
%
%
%
\IEEEPARstart{E}{llipsometry} is a non-destructive and contactless linear-optical spectroscopic experimental technique, which, if properly calibrated and performed correctly, can provide numerous physical parameters of materials with high precision and high accuracy~\cite{Fujiwara_2007}. The superiority of ellipsometry compared to pure intensity based techniques such as reflectance or transmittance measurements originates from the fact that the detected light is characterized in its entire polarization state of which intensity is only one parameter. Often, the polarization parameters are normalized to the intensity parameter which makes ellipsometry inherently less sensitive to spurious and almost always unavoidable variations of light sources, instrument apertures and light transfer functions. By determining phase and amplitude ratio of light reflected or transmitted upon samples, ellipsometry determines at least two independent quantities, which can be converted to the real and imaginary part of the dielectric function of the sample without the need to perform a numerical Kramers-Kronig analysis~\cite{Fujiwara_2007,Dressel_2002}.

Ellipsometry is a well-established technique from the infrared to the deep-ultraviolet spectral regions with broad and numerous scientific and commercial applications. Analysis of spectroscopic (generalized~\cite{Schubert96}) ellipsometry data can determine the thicknesses and isotropic (anisotropic) dielectric functions of individual layers within multiple-layered samples. The dielectric function of a material provides access to quantitative determination for many fundamental physical properties, for example, static and high-frequency dielectric permittivity, phonon modes, free charge carrier properties, band-to-band transition energies and life time broadening parameters. Expanding the accessible spectral range of spectroscopic ellipsometry to longer wavelengths, namely into the terahertz (THz) region, provides access to information about soft modes~\cite{PhysRevLett.101.167402}, DNA methylation~\cite{Cheon2016}, spin oscillations~\cite{BaierlS.2016}, antiferromagnetic resonances~\cite{doi:10.1063/1.4890475}, ferroelectric domains~\cite{doi:10.1063/1.4890939} and free charge carrier oscillations~\cite{Hofmannpss205_2008}, for example.


THz spectroscopic techniques are often performed as time-domain spectroscopy (TDS) or frequency-domain spectroscopy (FDS). THz TDS probes the response of materials to short-duration broad-frequency-band THz pulses created by using ultra-short laser switches. The signal is recored as a function of pump-delay time and a Fourier transformation converts the measured time-domain response into the frequency-domain response. THz TDS permits convenient access to the concept of sample modulation spectroscopy, where additional pump-probe instrumentation can be used to study the dynamic response of a sample to a cyclic disturbance. The THz probe beam field created by the ultra-short laser switch is typically characterized by low field energy, low spatial and temporal coherence, and therefore permits limited spectral resolution only. The low field intensities in THz TDS stem from a maximum optical-to-THz conversion efficiency between $10^{-3}$ and $10^{-4}$~\cite{Hoffmann:07,doi:10.1063/1.4733476}, resulting in a typical power below 100~$\mu$W, which is distributed over the THz spectrum~\cite{doi:10.1063/1.1459095,Globisch:15}. As a consequence THz TDS measurements that require optical windows or small apertures, such as measurements at low temperatures or high magnetic fields, are hampered due to inevitable intensity loss by transmission through multiple windows. Furthermore, the coherent length $l_c$ of THz pulses in THz TDS is in the order of a few millimeters~\cite{doi:10.1063/1.117511,Takayanagi:08,Tomasino2013}. Already for simple sample systems such as, for example, silicon substrates with a typical thicknesses of $d \approx 300-700~\mu$m and an index of refraction of $n \approx 3.4$, the front side reflected and the first order backside reflected THz beam only partially overlap coherently, and all higher order beams contribute by incoherent superposition. Correct computational modeling of this partial coherence beam overlay effect is difficult and requires accurate knowledge of the wavelength-dependent intensity distribution across the beam which probes the sample. Often, THz TDS raw data are therefore  modified (for example, data is cut out from a measured time delay spectrum towards larger delay times to remove echo pulses~\cite{ZHOU20161129}) in the attempt to remove the effect of substrate thickness interference. This modification, however, limits the spectral resolution. An inherent property of Fourier transform spectroscopy techniques, the spectral resolution $\Delta\nu$ in a THz TDS experiment depends on the length of the delay line $L$ and is given by $\Delta\nu=c/2L$, where $c$ is the speed of light. Typical THz TDS systems, including commercially available systems, have delay lines with $L\approx 5$~cm and therefore a typical maximum resolution of $\Delta\nu\approx 2$~GHz~\cite{Aenchbacher:10,Withayachumnankul2014}. Another challenge using a delay line is the precise positioning of its mirror. Positioning errors directly convert to errors in the time axis, therefore in the measured electrical fields $\mathbf{E}$, and thus, through data analysis, into an error of the complex refractive index~\cite{Jepsen:05,Jahn2016}. In THz TDS typically a reference signal is used to calculate the complex transmission function (attenuation and phase). This is only possible for transmission type measurements. For reflection type measurements self-referencing techniques such as ellipsometry must be employed, where self-referencing is often obtained by polarization modulation. While contemporary THz TDS ellipsometry equipment was demonstrated as a useful tool for materials characterization~\cite{doi:10.1063/1.1426258,MatsumotoJJAP48_2009,Neshat2012,doi:10.1063/1.4940976,doi:10.1063/1.4862974}, issues related to the low probe beam field intensities, the short coherence lengths and the limited spectral resolution remain.

THz FDS, and in particular THz FDS ellipsometry can overcome the limitations posed in THz TDS. Azarov \textit{et al.} reported on a FDS ellipsometer using the THz radiation emitted from a free electron laser as a light source~\cite{Azarov2015}.  Galuza~\textit{et al.} reported a FDS THz ellipsometer instrument using an avalanche-transit diode as light source~\cite{7416254}. While presenting the advantages of operating in the frequency domain such as high spectral resolution, these instruments require large-scale research facilities (free electron laser) or incorporate sources with large phase noise (avalanche-transit diode). Hofmann \textit{et al.}~\cite{hofmann_herzinger_schade_mross_woollam_schubert_2008} introduced the first backward wave oscillator (BWO) based ellipsometer. A BWO presents a continuously frequency tunable table top source for THz radiation with a high brilliance. BWOs emit THz radiation with high power reaching up to $1$~kW in military applications~\cite{kleen2012electronics}. Furthermore, BWOs possess a high phase stability and a small bandwidth~\cite{button2012infrared} and therefore a high coherence length. For example, the commercial BWO base tube used in the presented THz FDS ellipsometer, possesses a cw output power of 26~mW at 0.13~THz and a typical bandwidth of $\Delta f \approx 1$~MHz\footnote{Manufacturer specifications}. 
Hence, the intrinsic spectral resolution $\Delta\nu$ of the BWO tube is in the order of MHz ($\Delta\nu \geq \Delta f$). Employing the Wiener-Khinchin theorem~\cite{Wiener1930,Khintchine1934}, the coherence length is defined as the point at which the normalized Fourier-transform of the spectrum decays to $1/e$, which can be approximated as $l_c = c/\Delta f$ and where $c$ is the speed of light. For the used BWO this results in an intrinsic coherence length of $l_c \approx 300$~m for the emitted THz radiation. Though addressing issues posed by THz TDS, the usage of a THz radiation source with such a high coherence length requires strategies to suppress standing waves and to avoid speckle formation within the instrumentation as well as within the laboratory environment~\cite{Laser_Speckle_1975}.

In this paper, we present a BWO based THz FDS ellipsometer at the Terahertz Materials Analysis Center (THeMAC) at Link\"{o}ping university that follows concepts from stealth-technology. The instrument can incorporate a superconducting 8~T magnet and was designed to suppress speckle and standing wave formation by an enhanced \textit{geometry} (angeled walls, component inclination), \textit{coating} (absorbing and scattering surfaces) and \textit{modulation} (increased effective coherence length by frequency modulation). We discuss these design features and their effects on the spectral resolution and coherence length. Furthermore, we outline a method for THz frequency calibration. To demonstrate and benchmark the operation of the system, we present generalized ellipsometry data recorded in reflection and transmission over a wide range of angles of incidence for isotropic and anisotropic samples, as well as data from optical Hall effect, cavity enhancement and {\it in-situ} experiments from samples exposed to different gaseous environments.

\section{Ellipsometry: General Considerations}
\subsection{Formalisms}
\begin{figure}[htbp]
	\centering
	\includegraphics[
    width=0.35\textwidth]{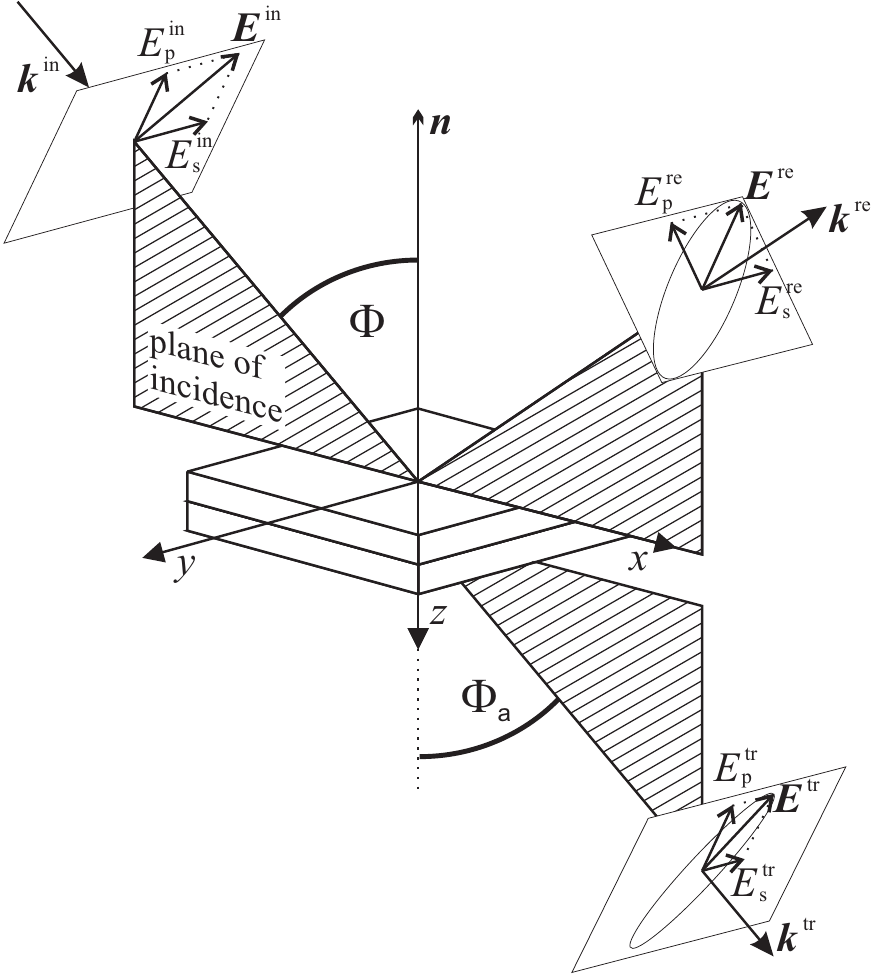}
	\caption{Schematic representation of the interraction of a plane electromagnetic wave with a sample. The wave vector $\bm{k}^{\text{in}}$ ($\bm{k}^{\text{re}}$, $\bm{k}^{\text{tr}}$) of the incoming (reflected, transmitted) electromagnetic plane wave and the sample normal $\bm{n}$ define the angle of incidence~$\bm{\mathrm{\varPhi}}$ (exit angle $\bm{\mathrm{\varPhi}}_{\text{a}}$) and the plane of incidence. The amplitudes of the electric fields of the incoming $\mathbf{E}^{\text{in}}$, reflected $\mathbf{E}^{\text{re}}$ and transmitted $\mathbf{E}^{\text{tr}}$ plane waves, can be decomposed into complex field amplitudes $E^{\text{in}}_{\text{p}}$, $E^{\text{in}}_{\text{s}}$, $E^{\text{re}}_{\text{p}}$, $E^{\text{re}}_{\text{s}}$, $E^{\text{tr}}_{\text{t}}$ and $E^{\text{tr}}_{\text{s}}$, where the indices p and s stand for parallel and perpendicular to the plane of incidence, respectively.}
	\label{fig:Schema_ellipsometry}
\end{figure}

The interaction of a plane electromagnetic wave with a sample (reflection or transmission) and the thereby caused change of polarization state (Fig.~\ref{fig:Schema_ellipsometry}) can be characterized using formalisms involving transition matrices. In ellipsometry, two different formalisms are often used, the Jones matrix and the Mueller matrix representation, which are based on electric fields and intensities, respectively. The Jones matrix $\mathbf{J}$ connects the electric field vectors before $\mathbf{E}^{\text{in}}$ and after interaction with the sample $\mathbf{E}^{\text{out}}$, while the Mueller matrix $\mathbf{M}$ connects the so called Stokes vectors before $\mathbf{S}^{\text{in}}$ and after interaction with the sample $\mathbf{S}^{\text{out}}$
\begin{equation}
	\mathbf{E}^{\text{out}}
	=
	\mathbf{J} \mathbf{E}^{\text{in}}
	\quad\quad\text{and}\quad\quad
	\mathbf{S}^{\text{out}}
	=
	\mathbf{M} \mathbf{S}^{\text{in}}\;.
\end{equation}
The Jones matrix $\mathbf{J}$~\cite{JONES41} is a dimensionless, complex-valued $2\times 2$ matrix, while the Mueller matrix $\mathbf{M}$ is a dimensionless, real-valued $4\times 4$ matrix
\begin{equation}
\label{eqn:def-jones}
	\mathbf{J}
	\hspace{-1pt}=\hspace{-2pt}
	\begin{pmatrix} j_{pp} & \hspace{-4pt}j_{ps} \\ j_{sp} & \hspace{-4pt}j_{ss} \end{pmatrix}
	\quad\hspace{-1pt}\text{and}\quad
	\mathbf{M}
	\hspace{-1pt}=\hspace{-2pt}
	\begin{pmatrix}
		M_{11} & \hspace{-4pt}M_{12} & \hspace{-4pt}M_{13} & \hspace{-4pt}M_{14} \\
		M_{21} & \hspace{-4pt}M_{22} & \hspace{-4pt}M_{23} & \hspace{-4pt}M_{24} \\
		M_{31} & \hspace{-4pt}M_{32} & \hspace{-4pt}M_{33} & \hspace{-4pt}M_{34} \\
		M_{41} & \hspace{-4pt}M_{42} & \hspace{-4pt}M_{43} & \hspace{-4pt}M_{44}
	\end{pmatrix}
	\!.
\end{equation}
The complex valued electric field vectors ($\mathbf{E}^{\text{in}}$ and $\mathbf{E}^{\text{out}}$) and the real valued Stokes vectors ($\mathbf{S}^{\text{in}}$ and $\mathbf{S}^{\text{out}}$) are in the p-s-coordinate system (see Fig.~\ref{fig:Schema_ellipsometry}) defined by
\begin{equation}
	\mathbf{E}
	\hspace{-1pt}=\hspace{-3pt}
	\begin{pmatrix}
		E_{p}\\
		E_{s}
	\end{pmatrix}
	\quad\hspace{-5pt}\text{and}\quad
	\mathbf{S}
	\hspace{-1pt}=\hspace{-3pt}
	\begin{pmatrix}
		I_{p} \hspace{-2pt}+\hspace{-2pt} I_{s} \\
		I_{p} \hspace{-2pt}-\hspace{-2pt} I_{s} \\
		I_{\hspace{-2pt}\botharrowm} \hspace{-2pt}-\hspace{-2pt} I_{\hspace{-5pt}\botharrowp\hspace{-1pt}} \\[-3pt]
		I_{\hspace{-1pt}R\hspace{-1pt}} \hspace{-2pt}-\hspace{-2pt} I_{\hspace{-1pt}L\hspace{-1pt}}
	\end{pmatrix}
	\hspace{-3pt}=\hspace{-3pt}
	\begin{pmatrix}
		E_{p}E^*_{p} \hspace{-2pt}+\hspace{-2pt} E_{s}E^*_{s} \\
		E_{p}E^*_{p} \hspace{-2pt}-\hspace{-2pt} E_{s}E^*_{s} \\
		E_{p}E^*_{s} \hspace{-2pt}+\hspace{-2pt} E_{s}E^*_{p} \\
		\hspace{-1pt}\text{i}(E_{p}E^*_{s} \hspace{-2pt}-\hspace{-2pt} E_{s}E^*_{p})\hspace{-3pt}
	\end{pmatrix}\!,
	\label{eqn:stokes_def}
\end{equation}
where $\text{i}=\sqrt{-1}$ is the imaginary unit, and the indices $p$, $s$, \hspace{-5pt}$\botharrowpnorm$, $\botharrowmnorm$, $R$ and $L$ on the electric fields $E$ and intensities $I$ denote linearly polarized parallel, perpendicular, $+45^{\circ}$ and $-45^{\circ}$ to the plane of incidence and right- and left-handed circularly polarized, respectively.

\subsection{Fresnel coefficients}
The four matrix elements $j_{mn}$ of the Jones matrix can be divided into reflection $r_{mn}$ and transmission $t_{mn}$ coefficients, and are in case of interaction with a single interface better known as the Fresnel coefficients for polarized light~\cite{Hecht_1987}. The individual coefficients are defined by
  \begin{subequations}
	\label{eqn:def-fresnel}
		\begin{align}
		&
		\begin{split}
			r_{pp}=
			\left.\left(
				\frac{E^{\text{re}}_{p}}{E^{\text{in}}_{p}}
			\right)\right|_{E^{\text{in}}_{s}=0}\;,
			\quad\quad
			r_{ps}
			&=
			\left.\left(
				\frac{E^{\text{re}}_{s}}{E^{\text{in}}_{p}}
			\right)\right|_{E^{\text{in}}_{s}=0}\;,
			\\
			r_{sp}=
			\left.\left(
				\frac{E^{\text{re}}_{p}}{E^{\text{in}}_{s}}
			\right)\right|_{E^{\text{in}}_{p}=0}\;,
			\quad\quad
			r_{ss}
			&=
			\left.\left(
				\frac{E^{\text{re}}_{s}}{E^{\text{in}}_{s}}
			\right)\right|_{E^{\text{in}}_{p}=0}\;,
		\end{split}\\
	&{~}\nonumber\\
	&
		\begin{split}
			t_{pp}=
			\left.\left(
				\frac{E^{\text{tr}}_{p}}{E^{\text{in}}_{p}}
			\right)\right|_{E^{\text{in}}_{s}=0}\;,
			\quad\quad
			t_{ps}
			&=
			\left.\left(
				\frac{E^{\text{tr}}_{s}}{E^{\text{in}}_{p}}
			\right)\right|_{E^{\text{in}}_{s}=0}\;,
			\\
			t_{sp}
			=
			\left.\left(
				\frac{E^{\text{tr}}_{p}}{E^{\text{in}}_{s}}
			\right)\right|_{E^{\text{in}}_{p}=0}\;,
			\quad\quad
			t_{ss}
			&=
			\left.\left(
				\frac{E^{\text{tr}}_{s}}{E^{\text{in}}_{s}}
			\right)\right|_{E^{\text{in}}_{p}=0}\;,
		\end{split}
		\end{align}
  \end{subequations}
where $E^{\text{in}}_{p}$, $E^{\text{in}}_{s}$, $E^{\text{re}}_{p}$, $E^{\text{re}}_{s}$, $E^{\text{tr}}_{p}$ and $E^{\text{tr}}_{s}$ are the projections of the electric field vectors parallel~($p$) and perpendicular~($s$) to the plane of incidence for the incoming (in), reflected (re) and transmitted (tr) wave.

\begin{figure*}[t]
	\centering
	\includegraphics[
    width=0.98\textwidth]{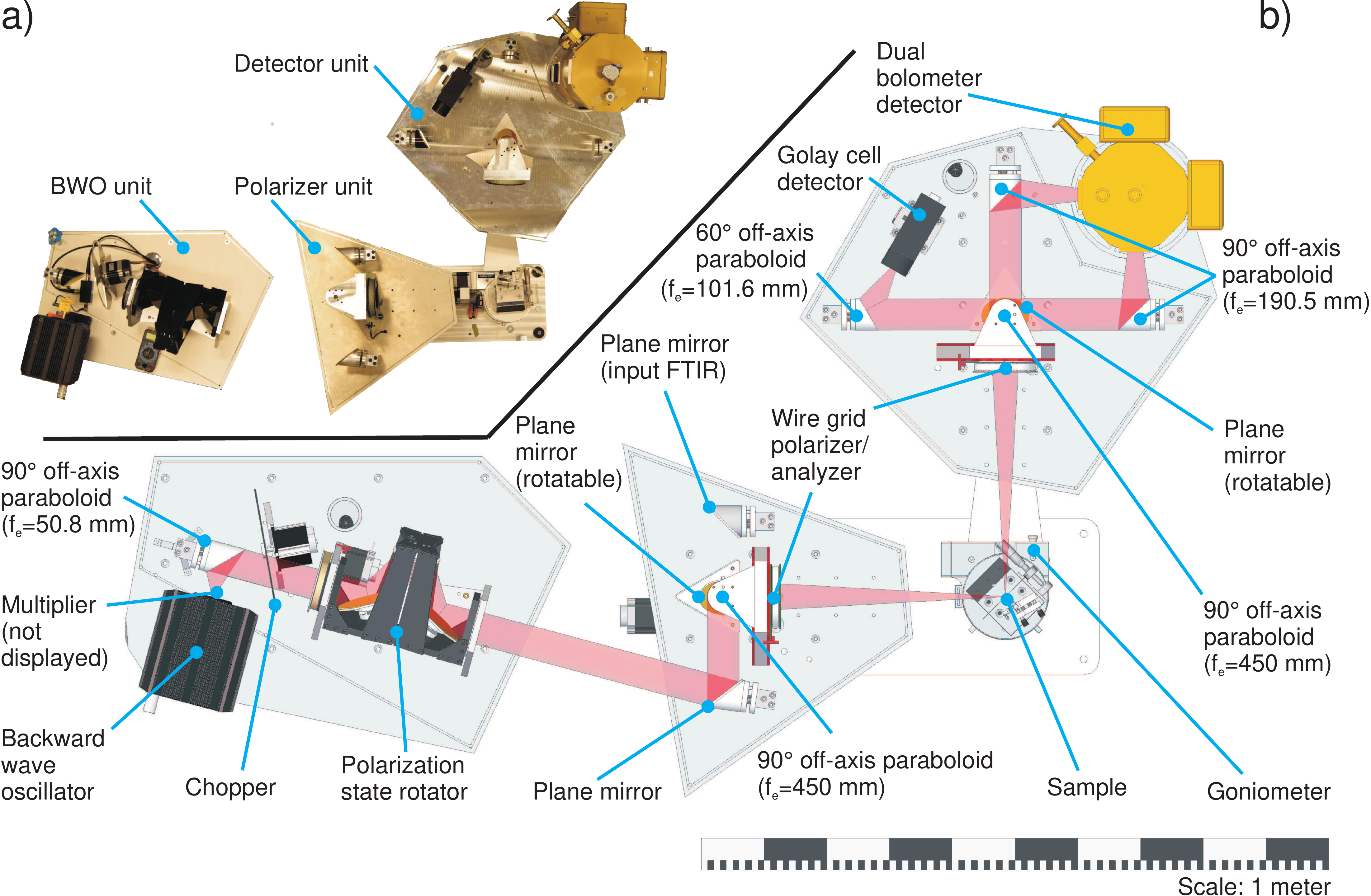}
	\caption{a) Photograph (top view) of the THz FDS ellipsometer. b) Technical drawing (top view) of the THz FDS ellipsometer with major components indicated and without absorbing foam sheets and housing. Note the angled outer geometry and the angled arrangement of components, resembling the concepts of stealth technology, employed to reduce and avoid standing wave formation at high resolution conditions.}
	\label{fig:CAD_top_view}
\end{figure*}

\subsection{Standard ellipsometry}
Standard ellipsometry (SE) is based on a special case of the Jones formalism and can be applied in case of isotropic non-depolarizing samples. In such case the cross-polarizing elements $j_{ps}$ and $j_{sp}$ of the Jones matrix can be neglected. SE determines the ratio between $j_{p}$ and $j_{s}$ (second index is dropped) and expresses this ratio $\rho$ by the ellipsometric parameters $\Psi$~and~$\Delta$
\begin{equation}
	\rho=\frac{j_p}{j_s}=\tan\Psi\exp\text{i}\Delta\;.
	\label{eqn:std-ellips}
\end{equation}
Note that $j_{m}=r_{m}$ for reflection type ellipsometry and $j_{m}=t_{m}$ for transmission type ellipsometry.

\subsection{Generalized ellipsometry}
\paragraph{Jones matrix ellipsometry}
For non-depolarizing samples Jones matrix generalized ellipsometry provides a unique quantification and a complete description of the change in the polarization state\footnote{This definition is convenient for rotating analyzer ellipsometers~\cite{SchubertIRSEBook_2004}. An alternative definition for Eqn.~(\ref{eqn:def-GE-Jones-b}) is $\rho_{ps}'=\frac{r_{ps}}{r_{ss}}=\tan\Psi_{ps}' \exp{(\text{i}\Delta_{ps}')} = \rho_{ps} \rho_{pp}$~\cite{SchubertJOSAA13_1996}}
\begin{subequations}
\label{eqn:def-GE-Jones}
\begin{align}
	\rho_{pp}&=\frac{r_{pp}}{r_{ss}}=\tan{\Psi_{pp}}\exp(\text{i}\Delta_{pp})\;,\\
	\rho_{ps}&=\frac{r_{ps}}{r_{pp}}=\tan{\Psi_{ps}}\exp(\text{i}\Delta_{ps})\;,\label{eqn:def-GE-Jones-b}\\
	\rho_{sp}&=\frac{r_{sp}}{r_{ss}}=\tan{\Psi_{sp}}\exp(\text{i}\Delta_{sp})\;.
\end{align}
\end{subequations}
Note that the Jones matrix [Eqn.~(\ref{eqn:def-jones})] contains four complex values, and therefore eight pieces of information~\cite{goldstein2011polarized}. However, the absolute phase of the $p$- and $s$-polarized components of the electromagnetic wave, and their amplitude (intensity) can be neglected. Therefore, the ellipsometric parameters $\Psi_{pp}$, $\Delta_{pp}$, $\Psi_{ps}$, $\Delta_{ps}$, $\Psi_{sp}$ and $\Delta_{sp}$ only provide six independent pieces of information.

\paragraph{Mueller matrix ellipsometry}
Mueller matrix generalized ellipsometry determines the Mueller matrix or parts of the Mueller matrix.
In contrast to Jones matrix generalized ellipsometry, Mueller matrix generalized ellipsometry provides a unique quantification and a complete description of the change in the polarization state even for depolarizing samples. The determined elements of the Mueller matrix are typically normalized by the Mueller matrix element $M_{11}$.

In case of an isotropic non-depolarizing sample, the Mueller matrix elements can be directly converted into the standard ellipsometry parameter $\Psi$
 and $\Delta$
\begin{equation}
\begin{aligned}
	M_{12} = M_{21} &=-\cos 2 \Psi\\
             M_{22} &= 1\\
	-M_{43} =M_{34} &= \sin 2 \Psi \sin\Delta\\
	M_{33} = M_{44} &= \sin 2 \Psi \cos\Delta \;,
\end{aligned}
\label{eqn:std-MM-ellips}
\end{equation}
and all other Mueller matrix elements are zero.

%
%
%
%

\subsection{Data analysis}
Ellipsometry is an indirect method, which is used to determine the optical response of materials. Typically the optical response is quantified in terms of the complex index of refraction $N=n+\text{i} k$ or the dielectric function \mbox{$\varepsilon=\varepsilon_1+\text{i} \varepsilon_2=N^2$}. Only in case of reflection-type ellipsometry on an isotropic, semi infinite material a direct inversion of the parameters $\Psi$ and $\Delta$ is possible
\begin{equation}
	\varepsilon=\sin^2\varPhi\left[1+\tan^2\varPhi\left(\frac{1-\rho}{1+\rho}\right)^2\right]\;,
	\label{eqn:pseudo-eps}
\end{equation}
where $\rho$ is given by Eqn.~\ref{eqn:std-ellips}. For all other cases, e.g., for layered or anisotropic samples, stratified layer model calculation based on the $4\times 4$ matrix algorithm~\cite{Schubert96} have to be employed. During data analysis best-match model parameters are determined, for which the best match between experimental and calculated data is achieved by varying relevant model parameters~\cite{JellisonTSF313-314_1998}.

\section{Terahertz ellipsometer}
\subsection{Components}
The THz FDS ellipsometer discussed here operates in a rotating analyzer configuration (source--polarizer--sample--analyzer--detector) and is thus capable of measuring the upper-left $3\times 3$ block of the Mueller matrix~\cite{Fujiwara_2007}. Figure \ref{fig:CAD_top_view} a) and b) display a top-view photographic image and top-view technical drawing of the THz FDS ellipsometer, respectively, in which the major components of the instrument are indicated. The THz FDS ellipsometer can be divided into three units: (i) the BWO-, (ii) polarizer- and (iii) detector-unit. All three units can be housed (housing removed in Fig.~\ref{fig:CAD_top_view}), which allows to purge the system with dry air or nitrogen.

\begin{figure*}[htbp]
	\centering
	\includegraphics[
    width=0.98\textwidth]{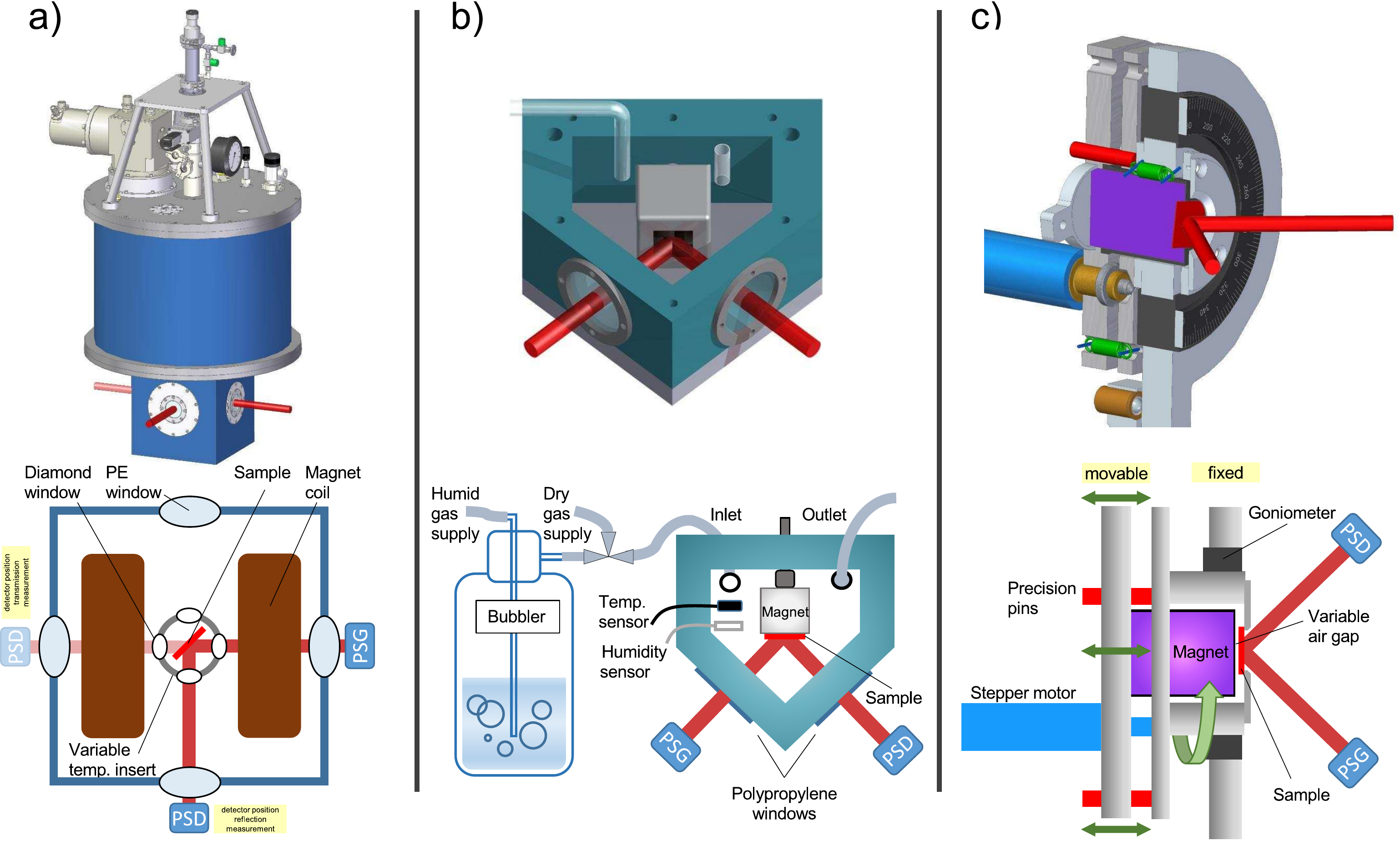}
	\caption{Technical (top) and schematic (bottom) drawings of three add-ons for the THz TDS ellipsometer: a) superconducting 8~T magnet with four optical ports (technical drawings printed with permission from Cryogenic Ltd, London UK), b) gas cell for {\it in-situ} measurements in different ambient conditions and c) air-gap-manipulator sample stage for cavity-enhanced optical Hall effect measurements. Major components are indicated in the schematic drawings together with the polarization state generator and detector (labeled: PSG and PSD).}
	\label{fig:addons}
\end{figure*}

\subsubsection{BWO unit}
The BWO unit contains a backward wave oscillator (BWO) model QS2-180 from SEMIC RF Electronic GmbH with the frequency range 97--179~GHz (base frequency range) as the source for THz radiation. The BWO unit has a power output of up to 26~mW and a high brilliance with a bandwidth of 1~MHz. Different GaAs Schottky diode frequency multipliers from Virginia Diodes, Inc. can be mounted to provide output frequencies of 202--352~GHz ($\times 2$: WR3.4x2 - Broadband Doubler, 6\% efficiency), 293--530~GHz ($\times 3$: WR2.2X3 - Broadband Tripler, 2\% efficiency), 670--1010~GHz ($\times 6$: WR3.4x2 - Broadband Doubler, 6\% efficiency and WR1.2X3 - Broadband Tripler, 1\% efficiency). In order to increase the antenna gain for the $\times6$ frequency multiplication, a waveguide matched horn antenna is mounted on the WR1.2X3 Broadband Tripler, improving the detector signal by a factor of 4. Due to the usage of waveguides, the emitted radiation is linearly polarized in the horizontal direction for all frequency multipliers. The THz radiation is then focused using a 90$^{\circ}$ off-axis paraboloid mirror with an effective focal length\footnote{The effective focal length $f_\text{e}$ is the distance between the focal point and the center of the off-axis paraboloid mirror. For 90$^{\circ}$ off-axis paraboloid mirror the effective focal length is twice the focal length $f_\text{e}=2f$.} of $f_e=50.8$~mm and a diameter of $50.8$~mm. The radiation then passes the 3 bladed chopper operated at 5.7~Hz, which leads to a chopping frequency of 17.0~Hz. The THz beam is then routed through the polarization state rotator, an odd bounce image rotation system~\cite{Herzinger_2004}, which revolves the polarization direction and is synchronized with the orientation of the polarizer in the polarizer unit. The THz radiation then leaves the BWO-unit and enters the polarizer unit.

\subsubsection{Polarizer unit}
In the polarizer unit the radiation is reflected of a plane mirror ($76.2$~mm diameter, gold sputtered) and reaches a rotatable plane mirror ($76.2$~mm diameter, gold sputtered), which is designed to switch between the BWO and a potential Fourier transform infrared (FTIR) spectrometer as the source of radiation. The beam is then focused by a gold coated 90$^{\circ}$ off-axis paraboloid mirror with an effective focal length of $f_e=450$~mm and a diameter of $50.8$~mm, and is guided through the polarizer towards the sample position. A free standing wire grid polarizer from Specac Ltd. with 5-$\mu$m-thick tungsten wires spaced 12.5~$\mu$m apart is used as polarizer, providing an extinction ratio
of approximately 3000, with 50~mm clear aperture and 70~mm outer diameter. The THz radiation leaves the polarizer unit to interact with the sample. The sample is mounted on a theta-2-theta goniometer, composed of a stack of two goniometers model 410 from Huber Diffraktionstechnik GmbH \& Co. KG, providing an angle of incidence~$\varPhi$ between 38$^{\circ}$ and 90$^{\circ}$ (0$^{\circ}$ and 90$^{\circ}$) for reflection (transmission) type experiments. 

\subsubsection{Detector unit}
After reflection from or transmission through the sample, the THz beam is guided into the detector unit and is routed through the analyzer. The analyzer rotates at 0.8~Hz and is equipped with the same type of free standing tungsten wire grid polarizer used as polarizer. The radiation then reaches a gold coated 90$^{\circ}$ off-axis paraboloid mirror with $50.8$~mm diameter and an effective focal length of $f_e=450$~mm. The collimated beam is reflected of a rotatable plane mirror ($76.2$~mm diameter, gold sputtered), which is designed to switch between three off-axis paraboloid mirrors  -- detector pairs. The two bolometer detectors (Infrared Laboratories, Inc.) are housed inside a common vacuum vessel and are each paired with a gold coated 90$^{\circ}$ off-axis paraboloid mirror with $50.8$~mm diameter and an effective focal length of $f_e=190.5$~mm. The vessel houses one general purpose 4.2~K bolometer and a 1.5~K bolometer optimized for the far-infrared spectral range. The golay cell is a room temperature detector and is paired with a gold coated 60$^{\circ}$ off-axis paraboloid mirror with $50.8$~mm diameter and an effective focal length of $f_e=101.6$~mm.

\subsubsection{System add-ons}
The THz FDS ellipsometer is designed to incorporate a superconducting 8~T magnet with four optical ports allowing reflection and transmission measurements (Fig.~\ref{fig:addons} a) with permission from Cryogenic Ltd, London UK), which can be moved in and out of the TDS ellipsometer using a rail system. During the construction of the THz FDS ellipsometer a minimum of ferromagnetic materials was used, and moving components, e.g., the rotating analyzer (see below) were manufactured from non-conducting materials (e.g., Acetal Copolymer) in order to avoid eddy currents.  

Figure~\ref{fig:addons} b) depicts a gas cell for environmental control made from Delrin and acrylic, with optical windows from 0.27~mm thick homopolymer polypropylene. The gas flow rate can be varied and has a maximum of approximately 0.5 liters/minute. The gas cell can be flushed with dry and humidified air, nitrogen, helium or other non-reactive gases. Furthermore, the gas cell is equipped with a high-grade neodymium permanent magnet with a surface field of $B=0.55$~T Tesla for \textit{in-situ} optical Hall effect measurements (see Sec.\ref{sec:in-situ}).

Figure~\ref{fig:addons} c) depicts a newly designed air-gap-manipulator sample stage for cavity-enhanced and cavity-enhanced optical Hall effect measurements~\cite{Knight2015}. The air-gap manipulator sample stage allows for opening an air gap between a highly reflective (metal) surface and the backside of the sample. The air gap thickness can be varied between 0 and 6~mm with an accuracy of 3~$\mu$m. The sample can be freely rotated in-plane and a permanent magnet with a surface field of up to $B=0.7$~T can be loaded into the air-gap manipulator. The air-gap manipulator sample stage allows for cavity-thickness scans and can, for example, help to decorrelate model parameters such as the free charge carrier effective mass and the sample (substrate) thickness during data analysis.

\begin{figure}[t]
	\centering
	\includegraphics[
    width=0.48\textwidth]{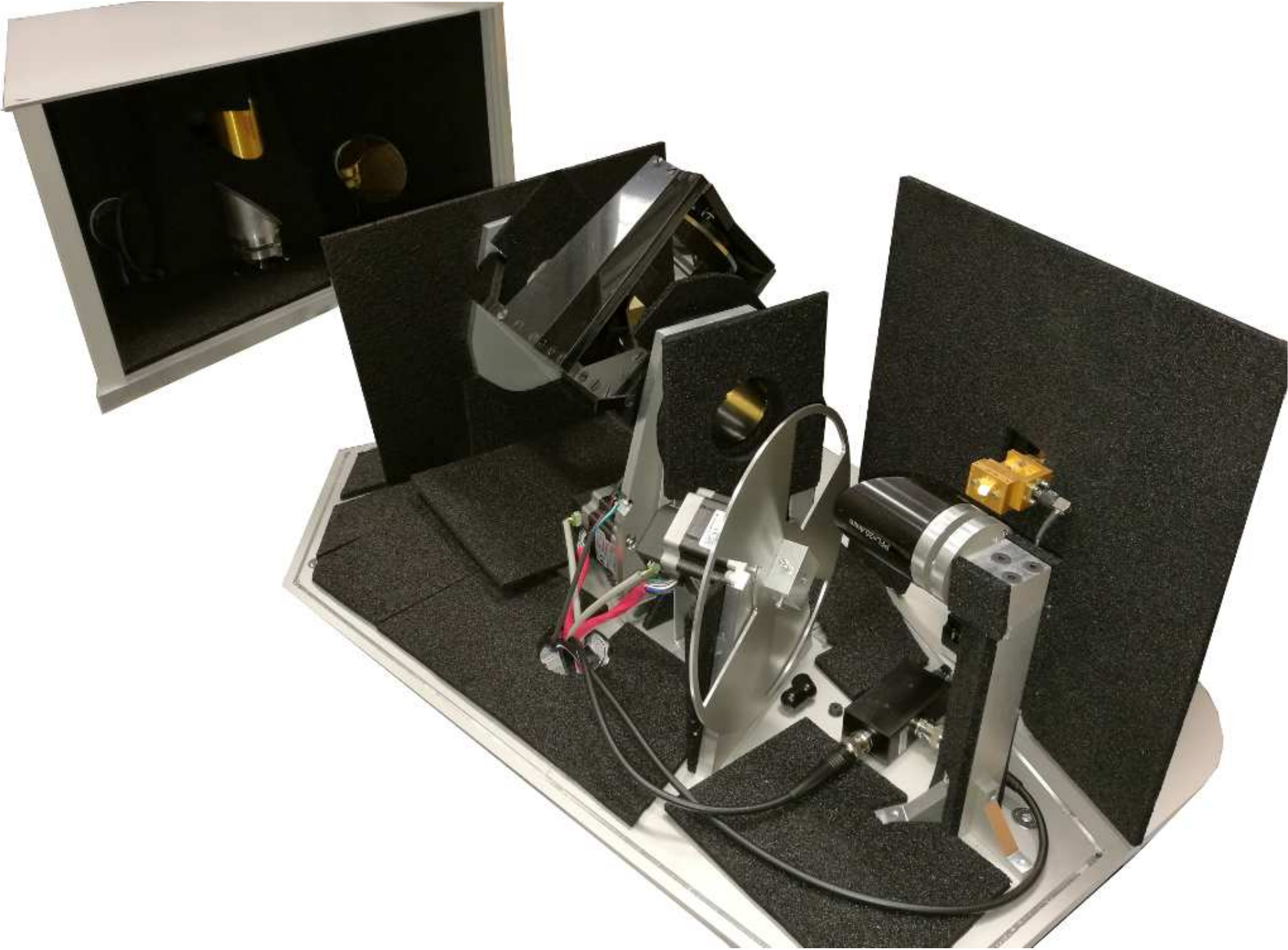}
	\caption{Photograph of the BWO unit (front, without housing) and the polarizer unit (back, with housing), with surfaces covered by THz radiation absorbing and scattering anti-static foam sheets (black material) and the angled components designed to suppress the formation of standing waves.}
	\label{fig:cond_foam}
\end{figure}

\subsection{Coherence / Stealth-technology based design}
Formation of standing waves pose challenges to the operation of a THz FDS ellipsometer based on BWO radiation sources. BWOs possess a high brilliance with narrow bandwidth, which results in a high coherence length. For example, for the BWO used in this work, the intrinsic bandwidth of the BWO tube is given as $\Delta \nu=1$~MHz by the manufacturer, which corresponds to a coherence length of approx. 300~m. Such a high value for the coherence length in combination with the fact that every metal component in the setup acts as a quasi-perfect mirror at THz frequencies, can generate standing waves patterns between different components (sample, BWO, detector, polarizers, framing, etc.). The problem about standing waves is twofold. For experiments requiring a source intensity baseline (e.g., simple transmission or reflection experiments), measurements with and without the sample (or with a reference standard) may generate different standing wave patterns in the experimental setup. Consequently, the intensities measured in the two experiments can be strongly affected, and therefore intensity normalization results in spurious effects. Further, for self-referencing techniques such as polarization modulation ellipsometry, standing waves between components of the experimental setup can result in a change of the polarization state detected by the instrument, and therefore can alter the quantities determined by an ellipsometric measurement. In both cases, for measurements in which standing waves appear, the experimental data can be affected by such standing wave patterns and as a result experimental data may not only represent the sample but rather the collective sample-instrument system. The THz FDS ellipsometer described here is designed to suppress standing waves, following concepts used in stealth technology. The strategy to suppress standing waves consists of three elements \textit{geometry}, \textit{coating} and \textit{modulation}.

The \textit{geometry} of the setup was, following principles from stealth technology, designed to contain a reduced number of surfaces and corners and to possess as few parallel surfaces as possible. These concepts were not only applied along propagation direction, but for all surfaces in the instrument due to scattering of forward propagating waves at apertures, entrances and exits (see Fig.~\ref{fig:CAD_top_view}). To further reduce effects from standing waves the chopper, polarizer/analyzer and structural elements of the sample holder are positioned under a 15$^{\circ}$, 10$^{\circ}$ and 15$^{\circ}$ angle with respect to the beam path, respectively.

The \textit{coating} consists of 6~mm thick porous anti-static foam sheets (Wolfgang Warmbier GmbH \& Co. KG Systeme gegen Elektrostatik), which possess a front (back) surface resistance of $R_s=10-100$~M$\Omega$ ($R_s=0.1-10$~M$\Omega$), a pore size of approximately $0.3-0.5$~mm and a void fraction of approximatively 50\%. The anti-static foam serves two purposes, due to its finite conductivity it acts as an absorber, and because the pore size is in the order of the wavelength it scatters and depolarizes THz radiation. Figure~\ref{fig:cond_foam} shows the anti-static foam in the BWO- and polarizer unit, where as many surfaces as possible were covered. This principle was applied throughout the whole setup, including the detector unit, the sample holder and the inside of the housing of the THz FDS ellipsometer. Independent from existing surfaces additional anti-static foam screens were added along the beam path to further suppress standing waves.

\begin{figure*}[ht]
	\centering
	\includegraphics[
    width=0.98\textwidth]{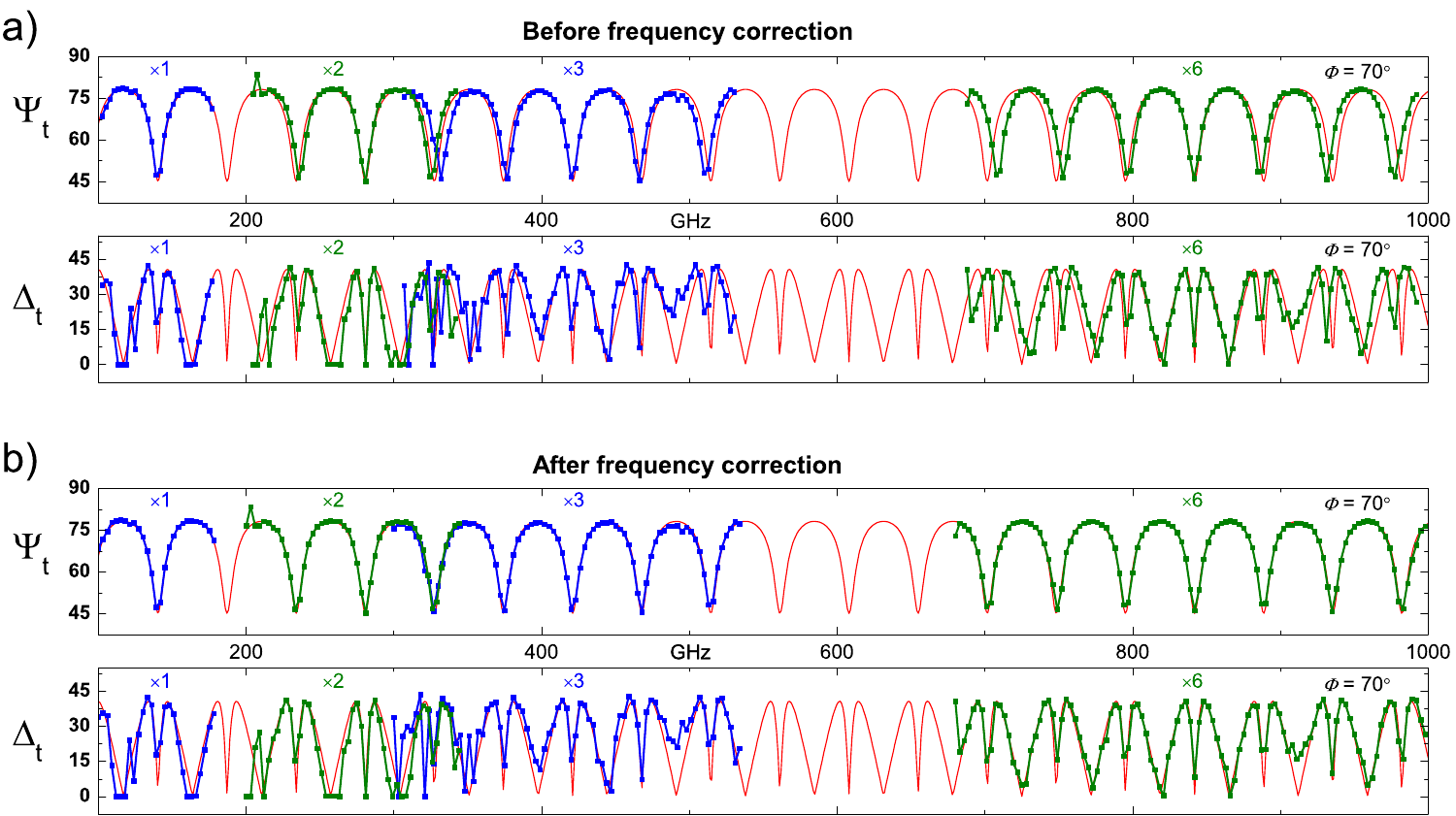}
	\caption{Experimental (symbols) standard ellipsometry transmission data $\Psi_t$, $\Delta_t$ and best model calculated (lines) data from various frequency multiplier ranges of a nominally 1~mm thick Silicon etalon for an angle of incidence $\Phi=70^{\circ}$ a) before and b) after the base tube $(\times 1)$ non-linear frequency calibration.}
	\label{fig:calib}
\end{figure*}

The \textit{modulation} of the BWO output frequency can be used to suppress effects from standing wave patterns. The BWO output frequency is determined and controlled by a DC-voltage power supply adjustable between $500$~V and $2600$~V. The DC BWO voltage can be superimposed by an additional sinusoidal 50~Hz voltage adjustable between $0$ and $25$~V in amplitude. The small voltage sinusoidal superposition leads to a temporal modulation of the BWO output frequency, which results in the increase of the effective frequency bandwidth of the continuous BWO radiation. Estimations based on calculations using the frequency-voltage calibration curve of the BWO for the $\times 6$ multiplication range and frequencies between 800~GHz and 850~GHz suggest an effective bandwidth of $\Delta\nu_{\text{eff}}$ of $0.5$~GHz/V$_{\text{mod}}$. Analysis of experimental data from \textit{c}-plane sapphire in the same spectral range, measured with different modulation voltages (0~V, 5~V, 10~V, 20~V and 50~V) results in an effective bandwidth of $0.45\pm 0.01$~GHz/V$_{\text{mod}}$. As a result the intrinsic coherence length, emitted by the BWO base tube ($\times 1$ multiplication) can be reduced from approx. 300~m at 0~V modulation, to an effective coherence length of approx. 30~cm at 25~V modulation.

\section{BWO source THz-Frequency calibration}
For the THz FDS ellipsometer the frequency of the THz radiation is determined by the high-voltage setting of the BWO cathode potential. The THz frequency vs. BWO cathode voltage function is in first order linear in voltage but requires non-linear voltage correction terms, and hence calibration. An example illustrating an insufficiently frequency calibrated BWO is shown in Fig.~\ref{fig:calib} a), which depicts experimental (symbols) standard ellipsometry transmission and respective best-match model calculated (lines) $\Psi_t$, $\Delta_t$ data for a nominally 1~mm thick silicon etalon at angle of incidence $\Phi=70^{\circ}$. A relatively good match between experiment and model is achieved for all frequency multipliers around a center frequency in each multiplication range, but a growing mismatch is observed towards the upper and lower ends of the respective multiplier range. Further, the magnitude of deviation scales with the frequency multiplication factor. The observed deviations between experiment and calculated data suggest a slight inaccuracy of the voltage-frequency calibration of the BWO ($\times 1$ multiplier) relation provided by the manufacturer. 

\begin{figure*}[t]
	\centering
	\includegraphics[
    width=0.98\textwidth]{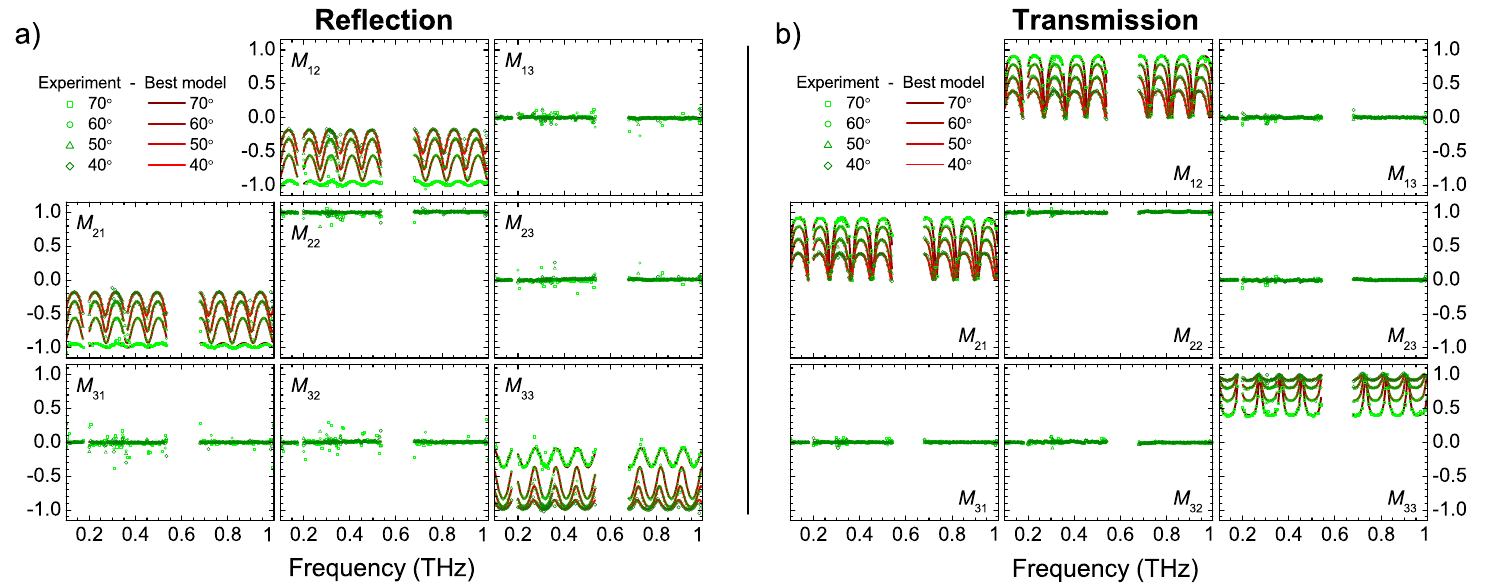}
	\caption{Experimental (symbols) and best-match model (lines) Mueller matrix data for a silicon wafer in reflection a) and transmission b) of a high-resistivity, floating-zone silicon wafer with a nominal thickness of $0.5$~mm at angles of incidence of $40^{\circ}$, $50^{\circ}$, $60^{\circ}$ and $70^{\circ}$.}
	\label{fig:trans_refl}
\end{figure*}

In order to frequency calibrate the THz FDS ellipsometer we augmented the method used by Naftaly \textit{et al.}~\cite{2010OptCo.283.1849N} to calibrate the frequency of THz TDS. The idea is to employ Fabry-P\'erot etalons with a neglectable absorption and dispersion of the index of refraction $(k=0,\;\frac{\partial n}{\partial \omega}=0)$ in the THz spectral range. Potential materials are, e.g., semi-insulating Si, semi-insulating 4H-SiC or sapphire.~\cite{1990JOSAB...7.2006G,2016OExpr..24.2590N}
%
We employed high-resistivity, floating-zone silicon because of its optical isotropy. A set of double-side polished Si wafers with a thickness of $d=0.5$~mm and 1~mm and a diameter of $101.6$~mm, cut out of the same ingot with nominal resistivity $\rho>3000~\Omega$\,cm was used.\footnote{For such resistivity values the free-charge careers make negligible contribution to the absorption coefficient and the index of refraction.} Additionally, a slab with a thickness of $d=3$~mm was created by stacking three cleaned segments of a 1~mm thick wafer together. Standard ellipsometry parameters $\Psi$ and $\Delta$ were recorded for all three thicknesses in transmission mode in all frequency ranges ($\times1$, $\times2$, $\times3$ and $\times6$ frequency multiplications) at four angles of incidence $\Phi=40^\circ$, $50^\circ$, $60^\circ$ and $70^\circ$.

The data was analyzed employing a linear regression algorithm (software: IDL) using an optical model describing the polarization response of an air/Si/air Fabry-P\'erot etalon with indexes of refraction \mbox{$n_{\text{air}}=1$} and $n_{\text{Si}}=n$ for air and Si ($k=0$ for both), respectively. The ellipsometric parameters $\Psi_t, \Delta_t$ can be written as
\begin{equation}
\begin{aligned}
	\tan\Psi_t\exp\text{i}\Delta_t &=\frac{t_s^2 \left(e^{\text{i} \alpha}-r_s^2\right)}{t_p^2 \left(e^{\text{i} \alpha}-r_p^2\right)}\;,
	\label{eqn:interference}
\end{aligned}
\end{equation}
where $t_p$ and $t_s$, $r_p$ and $r_s$ are the respective transmission and reflection Fresnel coefficients for $p$- and $s$-polarized light~\cite{Fujiwara_2007} for the air/etalon interface.
The parameter $\alpha=\frac{2 \omega d n}{c}\cos{\Phi}$ describes the total phase variation between the directly transmitted beam and the next higher transmitted beam, where $c$ and $\omega$ are the speed of light and the angular frequency, respectively. 

A 4th-order polynomial $f(U)$ for calibration of the output frequency vs. voltage of the BWO ($\times 1$ multiplication) was used
\begin{equation}
	f(U)=\sum^{4}_{i=0} a_i U^{i/2}\;.
	\label{eqn:frq_corr}
\end{equation}
Hence, the model parameters are the five coefficients of the frequency correction polynomial, the Si index of refraction and the wafer thicknesses. 
The thickness parameters of the nominally 0.5~mm and 1~mm thick wafers were determined from standard ellipsometry data recorded in reflection mode with an infrared ellipsometer (IRSE, J.A. Woollam Co., Inc.) at angles of incidence of 50$^\circ$, 60$^\circ$ and 70$^\circ$. This instrument provides ellipsometric data in the range 7.5--240~THz (250--8000~cm$^{-1}$).
The infrared ellipsometry data was analyzed assuming the optical constants of Si and their dispersions as determined in Ref.~\citenum{2013PSSBR.250..271S}, and we obtained
\mbox{$d=(494.86 \pm 0.01) \mu$m} and \mbox{$d=(975.71 \pm 0.01) \mu$m}. These parameters were kept constant during the THz frequency calibration procedure. The index of refraction of Si in the THz range was also kept constant at the literature value of $n=3.4172$~\cite{2013PSSBR.250..271S}.

Figure \ref{fig:calib}b shows best-match model calculated data where the non-linear frequency-voltage correction terms were included into the best-match model parameter base. As can be seen, the Fabry-P\'erot oscillations in the $\Psi$ and $\Delta$ data now match between experimental and model calculated data for the base tube and all frequency multiplication factors investigated here $(\times 2, \times 3, \times 6)$. The coefficients, as defined in Eqn.~\ref{eqn:frq_corr} are \mbox{$a_0=-18.6988$~GHz}, \mbox{$a_1=6.82632$~GHz/V$^{1/2}$}, \mbox{$a_2=-0.107034$~GHz/V}, \mbox{$a_3=1.33948\times 10^{-3}$~GHz/V$^{3/2}$} and \mbox{$a_4=-7.46583\times 10^{-6}$~GHz/V$^2$}.


\section{Applications}

\begin{figure}[p]
	\centering
	\includegraphics[
    width=0.46\textwidth]{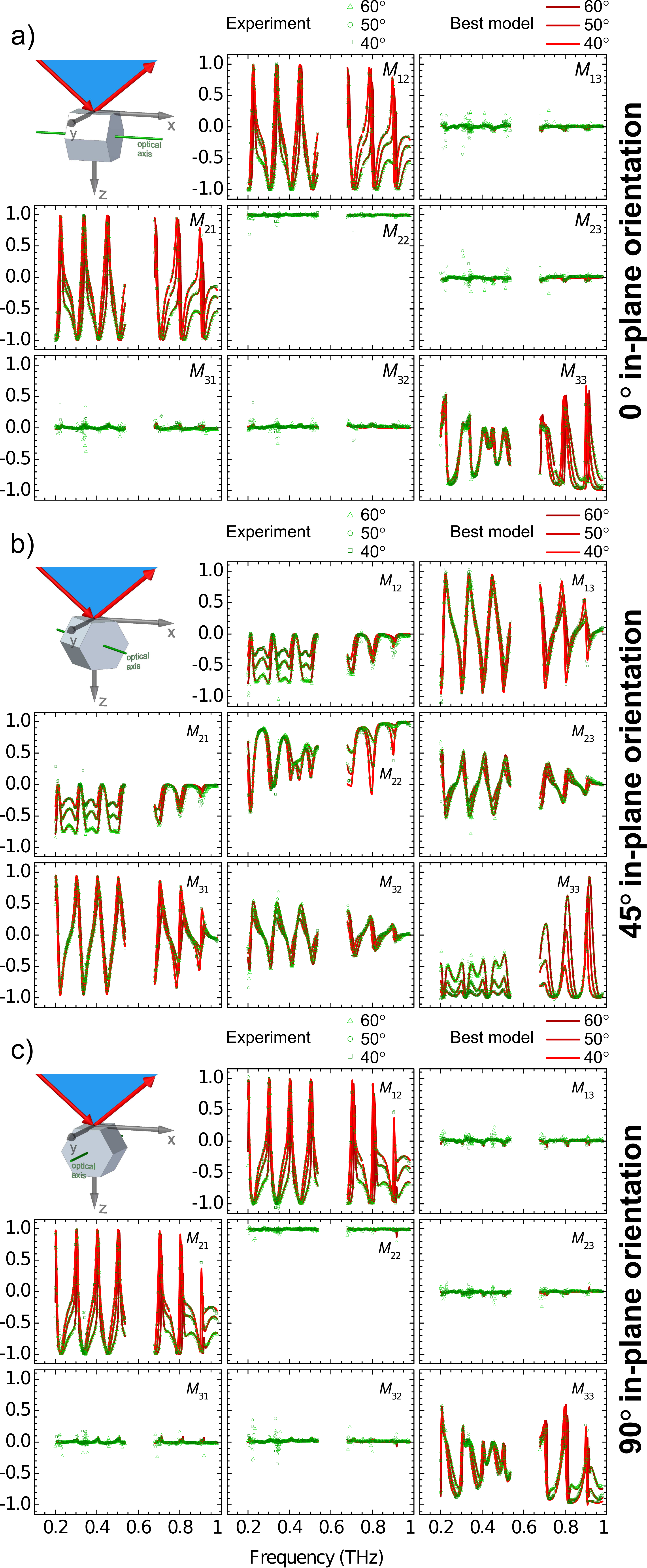}
	\caption{Experimental (symbols) and best-match model (lines) Mueller matrix data for m-plane $\alpha$-Al$_2$O$_3$ (sapphire) for three different in-plane orientations with azimuth angles a) $\Theta$ of 0$^{\circ}$, b) 45$^{\circ}$  and c) 90$^{\circ}$  between the optical axis of sapphire and the normal to the plane of incidence, at three angles of incidence (40$^{\circ}$, 50$^{\circ}$ and 60$^{\circ}$). The optical c-axis of the sapphire sample is indicated in the inserts.}
	\label{fig:m-plane}
\end{figure}

\subsection{Isotropic materials}
To demonstrate the reflection and transmission Mueller matrix capabilities of the THz FDS ellipsometer to isotropic materials we chose silicon. A high-resistivity, floating-zone silicon wafer different from the one used for the frequency calibration was measured. Figure~\ref{fig:trans_refl} displays the experimental (green) reflection and transmission Mueller matrix data, recorded at four different angles of incidence $\Phi=40^{\circ}$, $50^{\circ}$, $60^{\circ}$ and $70^{\circ}$, over the full frequency range from 0.1 THz to 1 THz ($\times1$, $\times2$, $\times3$ and $\times6$ frequency multiplications), together with corresponding best-match model calculated data (red). Note that the measured elements of the off-diagonal blocks of the Mueller matrix
$
\text{
	\scriptsize
	$
		\begin{bmatrix}
			M_{13}&\hspace{-3pt}M_{14}\\
			M_{23}&\hspace{-3pt}M_{24}\\
		\end{bmatrix}
	$
	\normalsize
	and
	\scriptsize
	$
		\begin{bmatrix}
			M_{31}&\hspace{-3pt}M_{32}\\
			M_{41}&\hspace{-3pt}M_{42}\\
		\end{bmatrix}
	$
}
$
vanish, as expected for isotropic materials~\cite{Fujiwara_2007}. All data was analyzed simultaneously and the only varied model parameters were the index of refraction, the Si thickness and the offset to the angle of incidence, which is kept constant for all angles of incidence. The excellent agreement between experimental and calculated data further demonstrates  that the frequency calibration procedure corrects for the initially observed frequency deviations. The best-match calculated 
\mbox{$n=3.418 \pm 0.001$} of Si agrees very well (within the error bars) with the value of $n=3.4172$ \cite{2013PSSBR.250..271S} used in the frequency calibration. This indicates that an accuracy of the index of refraction of one thousandth can typically be achieved with the THz ellipsometer. The best-match calculated values for the Si thickness and the offset in the angle of incidence were determined to be  
\mbox{$d=(498.2 \pm 0.2) \mu$m} and
\mbox{$\delta\Phi=(-0.19 \pm 0.02)^{\circ}$}, respectively.

\subsection{Anisotropic materials}
Many technologically relevant materials either in the form of active layers or substrates are optically anisotropic~\cite{GURNETT200639}. Examples include wide band gap materials for optoelectronics and high-frequency and power applications such as GaN, Ga$_2$O$_3$, ZnO, CdTe and substrates, such as silicon carbide, quartz and sapphire. Here we demonstrate THz FDS generalized ellipsometry characterization of sapphire as an example. Sapphire is uniaxial with its optical axis along the [0001] direction (\textit{c}-axis)~\cite{JEMT:JEMT1060020309}. Figure~\ref{fig:m-plane} displays experimental (green) Mueller matrix data recorded in reflection mode for $\alpha$-Al$_2$O$_3$ (sapphire) with an \textit{m}-plane (1$\bar{1}$00) surface orientation, together with corresponding best-model calculations (red). Three nominal in-plane orientations $(\Theta=0^{\circ}$, $45^{\circ}$ and $90^{\circ})$\footnote{The azimuth angle $\Theta$ is the angle between the optical axis (indicated in in the inserts of Fig.~\ref{fig:m-plane}) and the normal of the plane of incidence.} and three angles of incidence $(\Phi=40^{\circ}$, $50^{\circ}$ and $60^{\circ})$ were measured using the $\times2$, $\times3$ and $\times6$ frequency multipliers. Note that the measured block off-diagonal Mueller matrix elements ($M_{13,31}$, $M_{23,32}$) vanish for the high symmetry orientations $\Theta=0^{\circ}$ and $90^{\circ}$. 
On the other hand for the $\Theta=45^{\circ}$ in-plane orientation $p-s$ mode conversion appears and therefore all Mueller matrix elements reveal spectra different from zero. 


The THz data was analyzed together with standard ellipsometry data from a \textit{c}-plane \mbox{$\alpha$--Al$_2$O$_3$} sample recorded in the mid-infrared spectral range (IRSE, J.A. Woollam Co., Inc.), using the model described in Ref.~\citenum{SchubertPRB61_2000}. 
 During data analysis the $\Theta$-angles of each in-plane orientation and frequency multiplier range were varied independently. All $\Theta$-angels were found to be within $1^{\circ}$ of their nominal value. Further, the out-of-plane and the in-plane indices of refraction were varied and determined as \mbox{$\sqrt{\varepsilon_{\infty, \parallel}}=n_{\parallel}=(3.406 \pm 0.001)$} and \mbox{$\sqrt{\varepsilon_{\infty, \perp}}=n_{\perp}=(3.064 \pm 0.001)$}, respectively. These values are in excellent agreement with the refractive indices \mbox{$n_{\parallel}=3.407$} and \mbox{$n_{\perp}=3.067$} determined by far-infrared and THz TDS ~\cite{1990JOSAB...7.2006G}. The best model parameters for the sapphire thickness and angle of incidence offset were determined to be \mbox{$d=(445.9 \pm 0.1) \mu$m} and \mbox{$\delta\Phi=(-0.20 \pm 0.08)^{\circ}$}, respectively.

\begin{figure}[t]
	\centering
	\includegraphics[
    width=0.48\textwidth]{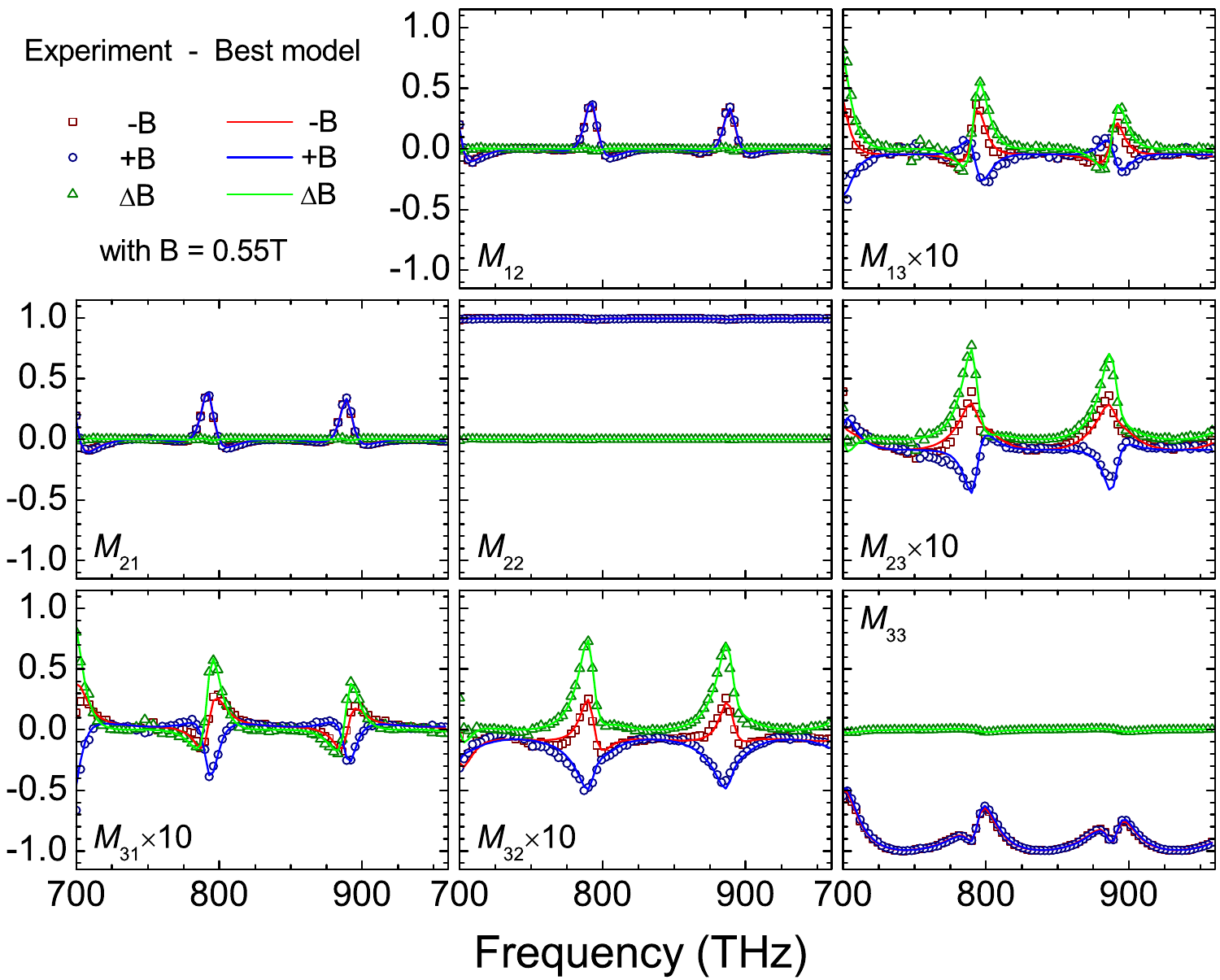}
	\caption{Experimental (symbols) and best model calculated (lines) cavity enhanced optical Hall effect data for AlGaN/GaN high electron mobility transistor structure (HEMT). The data was recorded with the magnetic ($B=\pm 0.55$~T) field parallel and antiparallel to the surface normal using a Neodymium permanent magnet. Green curves and scatter display the difference data between the two field directions.}
	\label{fig:HEMT}
\end{figure}

\subsection{Terahertz optical Hall effect (THz-OHE)}
The OHE is a physical effect analogue to the electric Hall effect, which appears at high (optical) frequencies in terms of optical birefringence and is measured using generalized ellipsometry ~\cite{Schubert:16}. The OHE allows for detection of three free charge carrier parameters, charge density, charge mobility and effective mass~\cite{KUHNE20112613, HofmannAPL_2012, SchoecheAPL103_2013, schoecheJAP_2013, Schoeche_JAP_2017}. The magnitude of the OHE birefringence can be further tuned by adding a cavity with varying thickness on the backside of a (transparent) sample~\cite{Knight2015, PSSC:PSSC201510214, ARMAKAVICIUS2017357}.

Here we present cavity-enhanced OHE in a high electron mobility transistor (HEMT) structure with a two dimensional electron gas (2DEG) channel as an example. We employ a neodymium permanent magnet with a surface field of $B=0.55$~T and a fixed cavity of $d=100 \mu\text{m}$ between the sample and the surface of the permanent magnet. The HEMT structure consists of  an Al$_{0.75}$Ga$_{0.25}$N barrier layer on 2-$\mu$m-thick GaN layer on a 4\textit{H}-SiC substrate employing an AlN nucleation layer~\cite{doi:10.1063/1.4922877}. Figure~\ref{fig:HEMT} shows experimental (symbols) OHE data together with best model calculations (lines) for two opposing magnetic field directions perpendicular to the sample surface, together with the respective difference. A model consisting of air/AlGaN//GaN 2DEG channel//GaN layer/AlN nucleation layer/SiC was used for the best-match model calculations. The model dielectric functions of all constituents included phonon contributions as described in Ref.~\cite{Schubert:16}. Only the 2DEG channel was found to exhibit free charge carriers, which was accounted for in the model dielectric function using the modified Drude model in the presence of magnetic field ~\cite{Schubert:16}. 

The OHE signal from the 2DEG is enhanced by the substrate/air-gap cavity and can be observed in the off diagonal block elements of the Mueller matrix $M_{13,23,31,32}$. Note that the symmetry properties of the corresponding off-diagonal elements of the Mueller matrix are fundamentally different for magneto-optic induced birefringence ($M_{13}=M_{31},M_{23}=M_{32}$, also $M_{22}=1$, see Fig.\,\ref{fig:HEMT}), compared to a birefringence induced by structural anisotropy as observed in \textit{m}-plane sapphire for azimuth angle   $\Theta$ of $45^{\circ}$  ($M_{13}=-M_{31},M_{23}=-M_{32}$, also $M_{22}\neq 1$, see Fig.~\ref{fig:m-plane}b). 
The best-match model parameters from the simultaneous analysis of data from all field directions and the difference data were a sheet carrier density of $n=(8.9 \pm 0.3)\times 10^{12} \text{cm}^{-2}$, effective mass of $m=(0.32 \pm 0.01)m_0$, and a mobility of $\mu=(1490 \pm 30)\text{cm}^{2}/ \text{Vs}$ for the 2DEG, $d=(501.20 \pm 0.03)\mu\text{m}$ for the 4\textit{H}-SiC substrate, and a air gap thickness of $d_{\text{gap}}=(103.5 \pm 0.3)\mu\text{m}$. The 2DEG density and mobility parameters are consistent with  the respective values typically reported for AlGaN/GaN HEMTs~\cite{doi:10.1063/1.3499656,doi:10.1063/1.4922877}. The 2DEG effective mass parameter is somewhat higher than  the bulk GaN electron mass of 0.232$m_0$~\cite{KasicPRB62_2000}, which is in agreement with previous reports~\cite{HofmannAPL_2012,doi:10.1063/1.4948582,PSSC:PSSC201510214}. The observed difference could be attributed to a penetration of the 2DEG wavefunction into the AlGaN barrier layer that was shown to be strongly affected by temperature~\cite{HofmannAPL_2012} and it is further discussed in Ref.~\cite{PSSC:PSSC201510214}.

\subsection{{\it in-situ} THz OHE}\label{sec:in-situ}

The FDS THz ellipsometer can be equipped with a gas cell for {\it in-situ} measurements under different ambient conditions (see Fig.~\ref{fig:addons}b).
As an example we present here a study of the effects of different, controlled gases and ambient humidity conditions onto the free charge carrier properties of epitaxial graphene.

\begin{figure}[bthp]
	\centering
	\includegraphics[
    width=0.49\textwidth
    ]{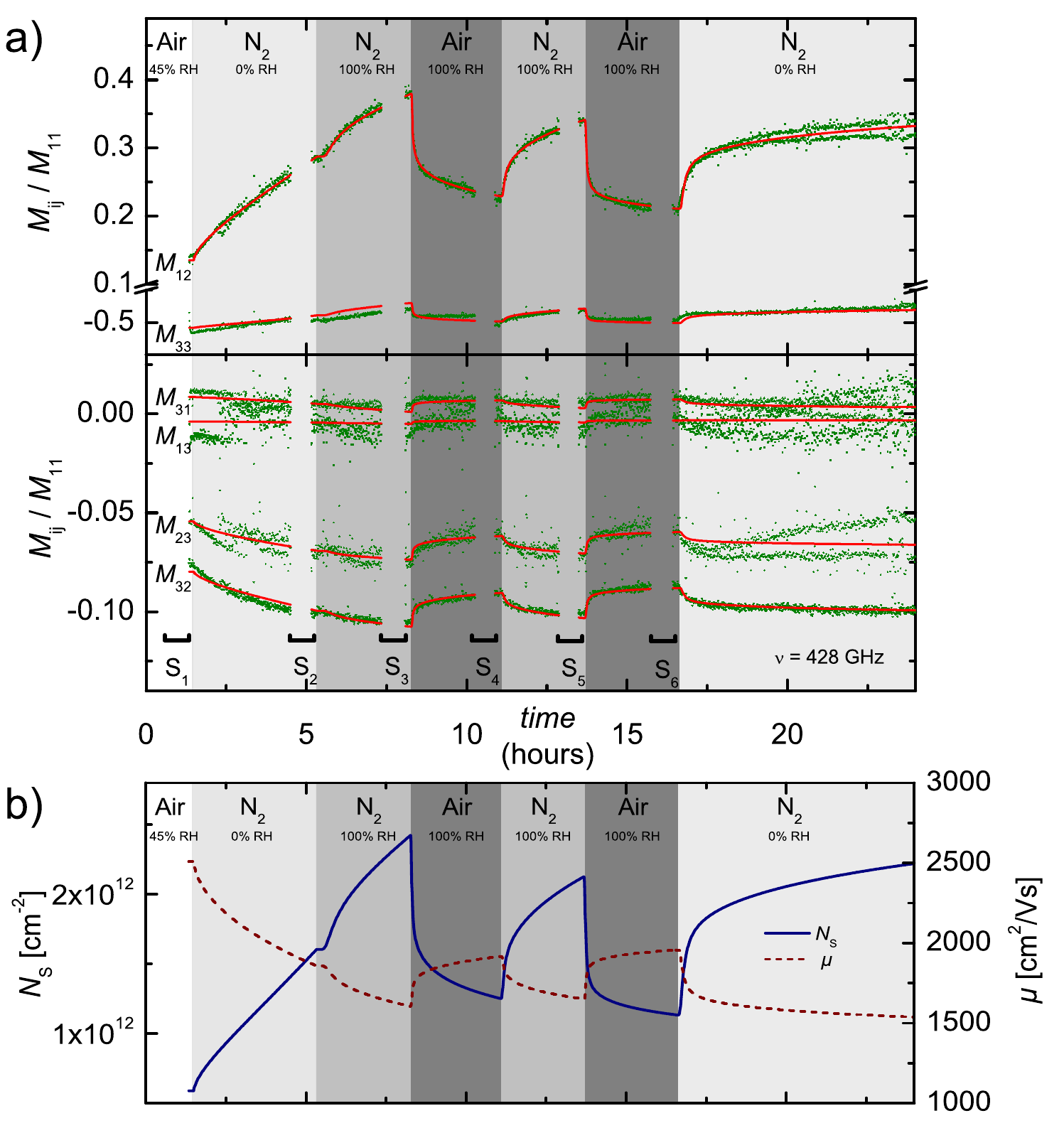}
	\caption{ a) Experimental (symbols) and best-matched model calculated (lines) in-situ OHE data for epitaxial graphene exposed cyclically to air and N$_2$ with different humidities. b) Best-match model free charge carrier density, $N_s$ and mobility, $\mu$ parameters of epitaxial graphene exposed to air and N$_2$ with different humidities.}
	\label{fig:dyn_graph}
\end{figure}

Graphene, like any other 2D material has an extremely large surface to volume ratio and its properties are therefore highly sensitive to the adjacent media (substrates, liquids, gases)~\cite{GaskillJAP111_2012,GiuscaAMI2_2015, NomaniSAB150_2010,YangAPL98_2011,ChanACSnano6_2012,DochertyNatureComm3_2012}. For a variety of growth methods (epitaxial, exfoliated, CVD grown) ambient doping can be observed~\cite{GaskillJAP111_2012,GiuscaAMI2_2015, NomaniSAB150_2010, YangAPL98_2011, ChanACSnano6_2012}. Here we present in-situ FDS THz OHE data from monolayer graphene epitaxially grown on 4\textit{H}-SiC in an Argon atmosphere. The sample was cyclically exposed to air and N$_2$ with different humidities. The experiment was conducted at a fixed THz frequency ($428$~GHz), at an angle of incidence of $\Phi=45^{\circ}$, using a neodymium permanent magnet with a surface field of $B=0.55$~T and a fixed cavity of $d=100$~$\mu$m between the sample and the surface of the permanent magnet. Figure~\ref{fig:dyn_graph}a shows {\it in-situ} OHE data with the sample exposed to different gas environments and relative humidities. 
For the data analysis an optical model consisting of air/graphene/SiC/air-gap/magnet was used. The model dielectric function of the semi-insulating SiC substrates contains only a phonon contribution, while the neodymium magnet is treated as a metal with a Drude dielectric function. Both dielectric functions of SiC and the magnet were determined in separate experiments and included in the data analysis without any changes.  The model dielectric function of graphene included contributions from free charge carriers in the presence of magnetic field as detailed in Ref.~\citenum{ARMAKAVICIUS2017357}. The effective mass of graphene was coupled to the sheet carrier concentration $N_s$ using $m^*=\sqrt{(h^2N_s)/(4\pi v_{\text{f}}^2)}$ from Ref.~\citenum{NovoselovN438_2005}, where $h$ and $v_{\text{f}}$ are the Plank constant and the Fermi velocity, respectively. The THz OHE showed that free charge carriers in the epitaxial graphene were electrons. The results on best-match calculated free electron density $N_s$ and mobility $\mu$  parameters of epitaxial graphene are shown in Fig.~\ref{fig:dyn_graph}b.

It can be seen that significant changes in the free charge carrier properties occur upon cyclic variation of the environment from air to N$_2$. We have recently reported similar findings for single layer graphene exposed to N$_2$ and He with 0\% relative humidity \cite{Knight2017}. Here, an increase of the electron density is observed, independent on the relative humidity, when N$_2$ is injected into the gas cell. The process can be reversed to a certain extent, and the carrier density can be reduced by injecting air with 100\% humidity. This behavior indicates that the ambient-doping mechanism depends on components in air other than N$_2$, namely O$_2$ or CO$_2$. A possible doping mechanism based on a redox reaction involving H$_2$O, O$_2$ and CO$_2$ is discussed in Ref.~\citenum{GaskillJAP111_2012}. While this reaction may not be the only possible chemical process, it is described as the most dominant~\cite{ChakrapaniScience318_2007}. The largest variation in $N_s$ from $5.9 \times 10^{11}$~cm$^{-2}$ to $2.4 \times 10^{12}$~cm$^{-2}$ is observed in the first two cycles with initial state of graphene being already exposed to ambient conditions for several months. It is worth mentioning that no complete reversal in $N_s$ is observed in the following cycles. These results indicate that the electronic properties of epitaxial graphene may change over the course of several days, if the graphene sample underwent a change of the ambient conditions. This is important to consider if epitaxial graphene is studied, since the results may be changing drastically over time.

The exposure of epitaxial graphene to different gases also results in significant changes in its free electron mobility parameter, $\mu_n$ (Fig.~\ref{fig:dyn_graph}b). The observed variation in mobility can be correlated to the change in free hole density, $\mu_n \sim$ 1/$N_s$. This indicates that it is possible that only the carrier density in the graphene changes as a result of gas exposure and where the observed change in mobility is only given by the $\mu~(N_s)$ dependence. However, scattering due to charge impurities could not be excluded. Further details discussing the scattering mechanisms in epitaxial graphene under exposure of different gases can be found in Ref.~ \cite{Knight2017}.

\section{Conclusion}
We have presented the newly designed Terahertz (THz) frequency-domain spectroscopy (FDS) ellipsometer at the Terahertz Material Analysis Center (THeMAC) at Link\"oping university and have demonstrated its application to a variety of technologically important materials and heterostructures. We have shown that employing concepts used in stealth technology for the instrument  geometry and scattering anti-static coating, and modulation of the backward wave oscillator (BWO) THz source allows for effective suppression of standing waves enabling accurate ellipsometry measurements with high spectral resolution (of the order of MHz). We further have demonstrated an etalon-based method for frequency calibration in THz FDS ellipsometry. 

The instrument can incorporate various sample compartments, such as a superconducting magnet, \textit{in-situ} gas cells or resonant sample cavities, for example. Reflection and transmission ellipsometry measurements over a wide range of angles of incidence for isotropic (Si) and anisotropic (sapphire) bulk samples have been presented and used for the determination of the material dielectric constants. We have further demonstrated results from cavity enhanced THz optical Hall effect experiments on an AlGaN/GaN high electron mobility transistor structure (HEMT), determining the free charge carrier density, mobility and effective mass parameters of the 2D electron gas (2DEG) at room temperature. Finally, we have shown through \textit{in-situ} experiments on epitaxial monolayer graphene exposed to different gases and humidities that THz FDS ellipsometry is capable of determining  free charge carrier properties and following their changes upon variation of ambient conditions in atomically thin layers. Therefore, THz FDS ellipsometry
has been shown to be an excellent tool for studying free charge carrier and optical properties of thin films,  device heterostructures and bulk materials, which are important for current and future electronic applications. Exciting perspectives of applying THz FDS ellipsometry for exploring low-energy excitation phenomena in condensed and soft matter, such as the vibrational, charge and spin transport properties of magnetic nanolaminates, polymers and hybrid structures for photovoltaics and organic electronics; and determination of THz optical constants and signatures of security and metamaterials are envisioned.


\section*{Acknowledgment}
The authors would like to acknowledge Tino Hofmann (University of North Carolina, Charlotte) for the discussions and input that helped bringing this project forward. Further we would like to thank Craig M. Herzinger (J.A. Woollam Co., Inc.) for support in ellipsometry software development and for numerous fruitful discussions. We thank Jr-Tai Chen (SweGaN AB and Link\"oping University) for growing and providing the HEMT sample, Jan \v Sik (On Semiconductor) for providing intrinsic Si wafers, and Rositsa Yakimova (Link\"oping University and Graphensic AB) for collaboration regarding graphene growth.  We acknowledge support from the Swedish Foundation for Strategic Research (SSF, grants No. FFL12-0181, RIF14-055), \AA Forsk (grant No. 13-318), the Swedish Research Council (VR Contracts 2013-5580 and 2016-00889), the Swedish Governmental Agency for Innovation Systems (VINNOVA grant No. 2011-03486), the Swedish Government Strategic Research Area in Materials Science on Functional Materials at Link\"oping University (Faculty Grant SFO Mat LiU No 2009 00971), the National Science Foundation through the Center for Nanohybrid Functional Materials (EPS-1004094), the Nebraska Materials Research Science and Engineering Center (DMR-1420645), and awards CMMI 1337856 and EAR 1521428.

\ifCLASSOPTIONcaptionsoff
  \newpage
\fi


%
%

%

\begin{IEEEbiography}[{\includegraphics[
width=1in,height=1.25in,clip,keepaspectratio]{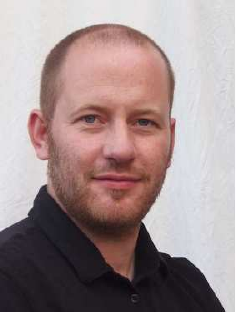}}]{Philipp K\"{u}hne}
received in 2009 the Dipl.-Phys. degree from the University of Leipzig, Leipzig, Germany. In 2014 he received a Ph.D. degrees in electrical engineering from the University of Nebraska, Lincoln, USA. Since 2014 he works as a post-doctoral researcher and research engineer at the Terahertz Materials Analysis Center (TeMAC) at the Link\"{o}ping University, Link\"{o}ping, Sweden. His work is focused on material analysis with emphasis on ellipsometry.
\end{IEEEbiography}

\begin{IEEEbiography}[{\includegraphics[
width=1in,height=1.25in,clip,keepaspectratio]{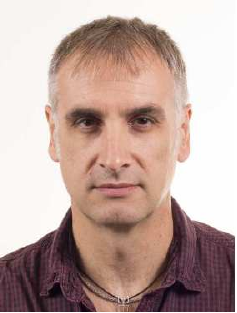}}]{Vallery Stanishev}
received a M.S. degree in astrophysics from Sofia University, Bulgaria and a Ph.D. degree in astrophysics from Institute of Astronomy, Bulgarian Academy of Sciences, in 1995 and 2002, respectively. He works as a research engineer and and university lecturer at the Link\"{o}ping University, Link\"{o}ping, Sweden. His research interests include  materials analysis with spectroscopic ellipsometry.
\end{IEEEbiography}

\begin{IEEEbiography}[{\includegraphics[
width=1in,height=1.25in,clip,keepaspectratio]{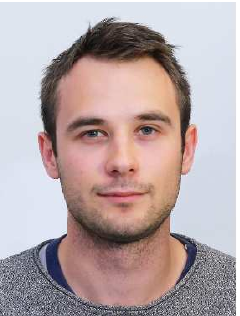}}]{Nerijus Armakavicius}
received B.S. and M.S. degrees in Applied Physics from Kaunas University of Technology, Lithuania, in 2012 and 2014, respectively. Since 2014 he works as a PhD student at the Terahertz Materials Analysis Center at the Link\"{o}ping University in Sweden. His research interests are spectroscopic ellipsometry, optical Hall effect, two-dimensional electronic materials and advanced semiconductors.
\end{IEEEbiography}

\begin{IEEEbiography}[{\includegraphics[
width=1in,height=1.25in,clip,keepaspectratio]{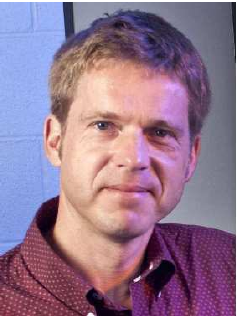}}]{Mathias Schubert}
received Dipl.-Phys., Dr. rer. nat., and Dr. habil. (Physics) degrees from the University of Leipzig, Leipzig, Germany, in 1994, 1997, and 2003, respectively. He received a Dr. tech. h.c. from Link\"{o}ping University, Sweden, in 2015. He became associate professor and full professor at the University of Nebraska, Lincoln, USA, in 2005 and 2012, respectively. He currently holds the J.A. Woollam Distinguished Professorship Chair. His research interests cover ellipsometry spectroscopy and imaging, the optical Hall effect, advanced semiconductors and functional materials, nanocomposites, optoelectronic devices, conductive organics, in-situ monitoring, and low-symmetry materials. He is also an Associate Editor of Applied Physics Letters.
\end{IEEEbiography}

\begin{IEEEbiography}[{\includegraphics[
width=1in,height=1.25in,clip,keepaspectratio]{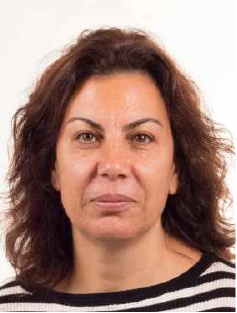}}]{Vanya Darakchieva}
received a Ph.D. degree in materials science and habilitation in semiconductor materials from the Link\"{o}ping University, Link\"{o}ping, Sweden in 2004 and 2009, respectively. She became an associate professor at Link\"{o}ping University, Link\"{o}ping, Sweden in 2012. She is head of the Terahertz Materials Analysis Center, and director of the Competence Center for III-Nitride Technology at Link\"{o}ping University. Her research is focused on novel semiconductor and nanoscale materials for high-frequency and power electronics, and advanced spectroscopic metrology of transport and electronic properties of materials.\blk
\end{IEEEbiography}



\end{document}